\documentclass[a4paper,11pt]{article}
\pdfoutput=1 
\usepackage{jinstpub} 
\usepackage{lineno}

\title{Advanced Radiation Detector Design for Applications in Food Safety and National Security}
\author[1]{A. Bross}
\author[2]{E.C. Dukes}
\author[1]{S. Hansen}
\author[1]{A. Pla-Dalmau}
\author[1]{P. Rubinov}
\affiliation[1]{Fermi National Accelerator Laboratory, Box 500, Batavia, IL USA}
\affiliation[2]{Physics Department, University of Virginia, Charlottesville, VA, USA}
\date{\today}

\abstract{
We describe a new concept for a radiation detector based on extruded scintillator technology and commercially available solid-state photo-detectors.  The detector is simple in construction, robust, very efficient, cost-effective and easily scalable in size from tens of cm$^2$ to tens of m$^2$.  We describe two possible applications: flagging radioactive food contamination and detection of illicit radioactive materials, such as those potentially used in a dirty bomb.
}


\arxivnumber{2307.04828} 
\def\be{\begin{equation}}
\def\ee{\end{equation}}
\def\bea{\begin{eqnarray}}
\def\eea{\end{eqnarray}}

\usepackage{caption}
\usepackage{lineno}
\usepackage{amsmath}
\usepackage{graphicx}
\usepackage{mathtools}
\usepackage{siunitx}
\usepackage{url}
\usepackage[toc,page]{appendix}
\begin{document}
\maketitle
%


%
%

%
%
%
%





\section{Overview}
\begin{figure}[t]
\centering 
\includegraphics[width=0.45\columnwidth]{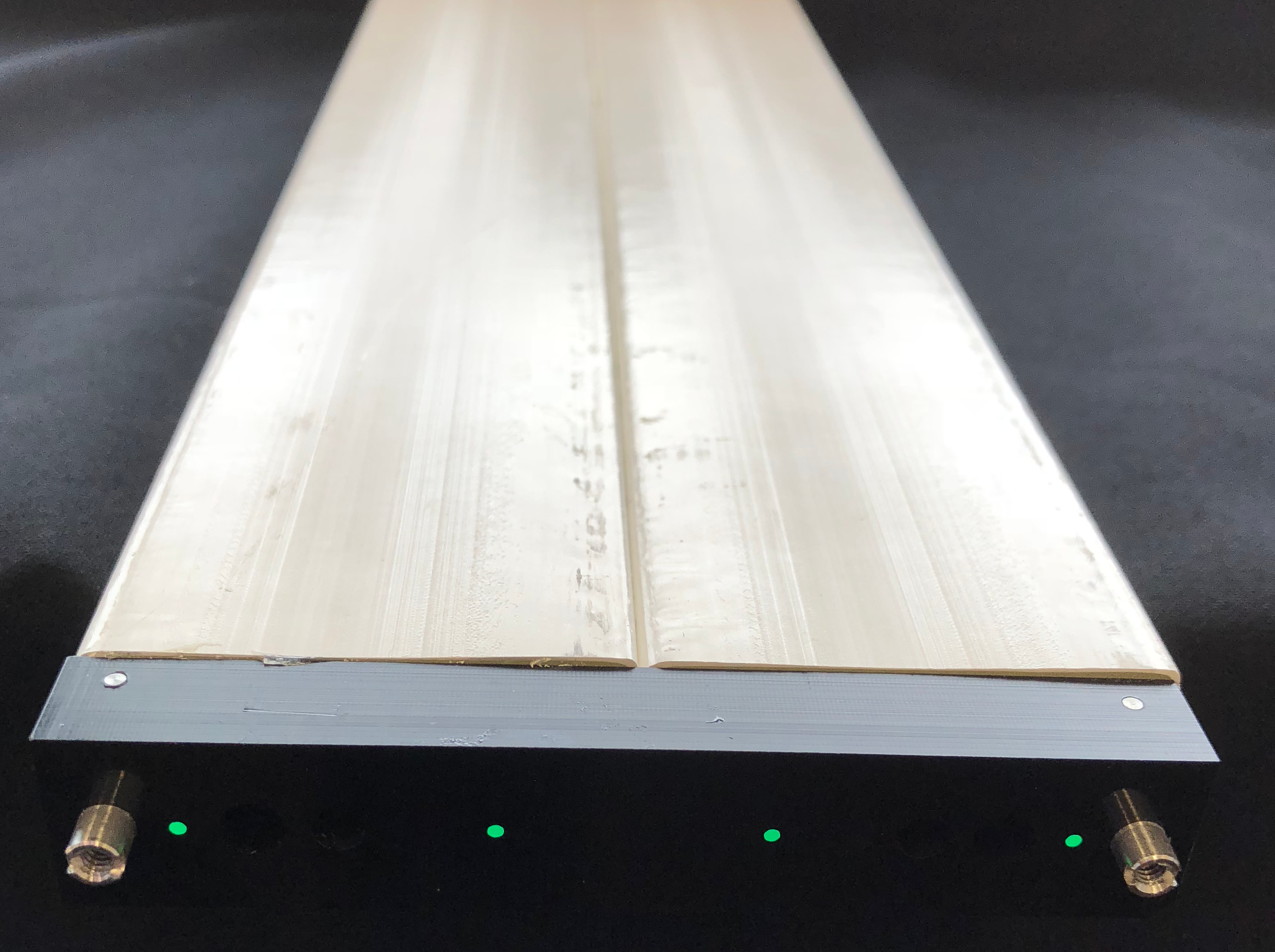}
\includegraphics[width=0.54\columnwidth]{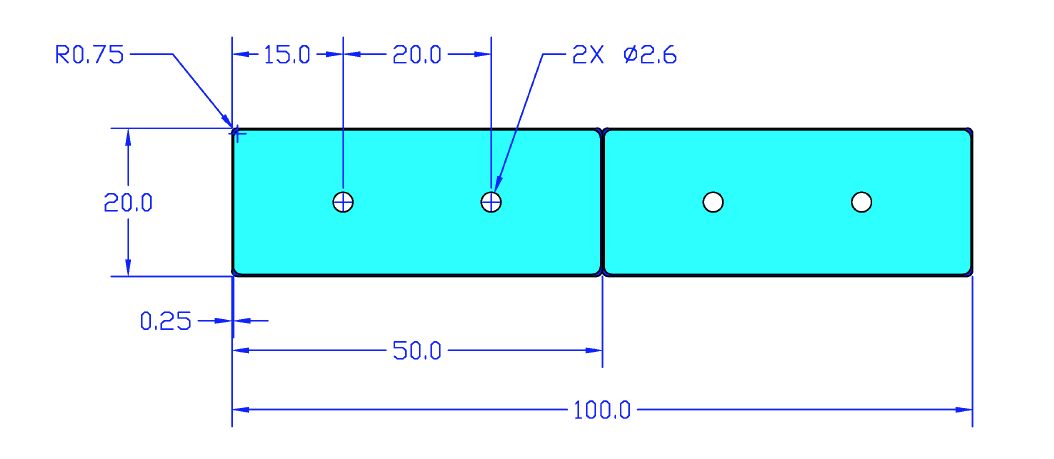} 
\caption{[Left]: Photograph of a Di-counter, [Right]: Nominal dimensions (mm) of a Di-counter.} 
\label{fig:dicounter}
\end{figure}

\begin{figure}[h]
\centering 
\includegraphics[width=0.45\columnwidth]{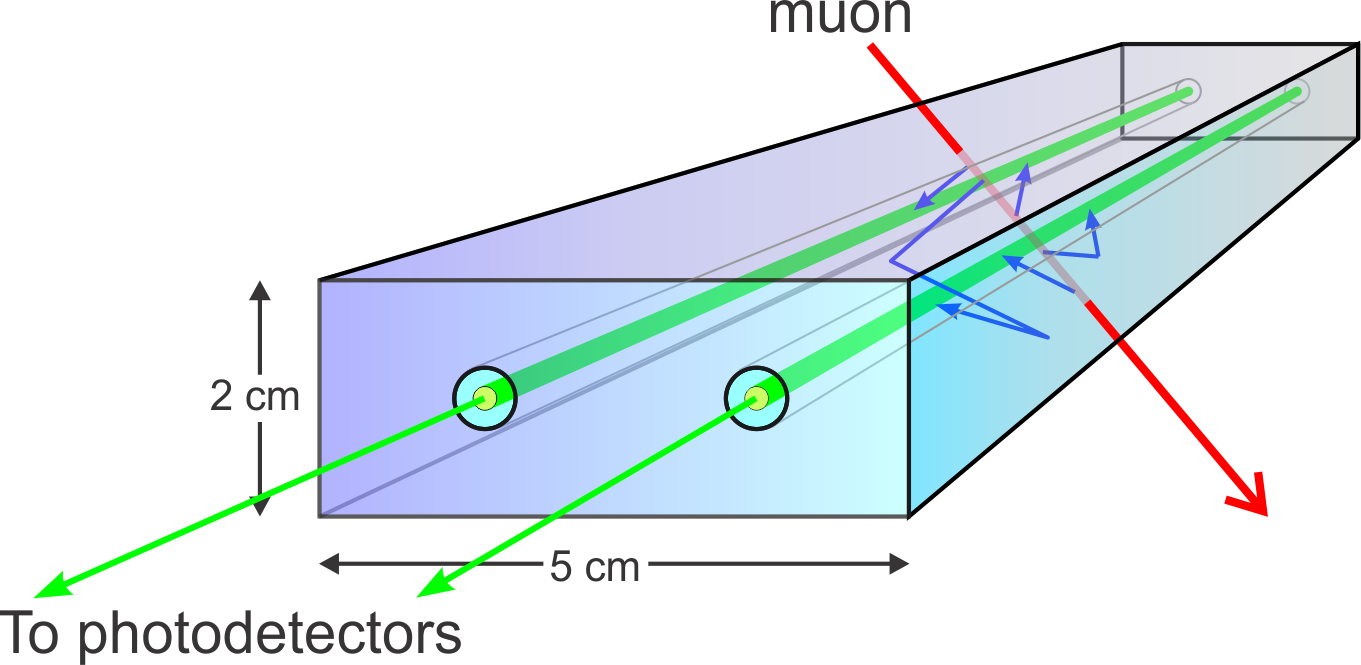}
\includegraphics[width=0.45\columnwidth]{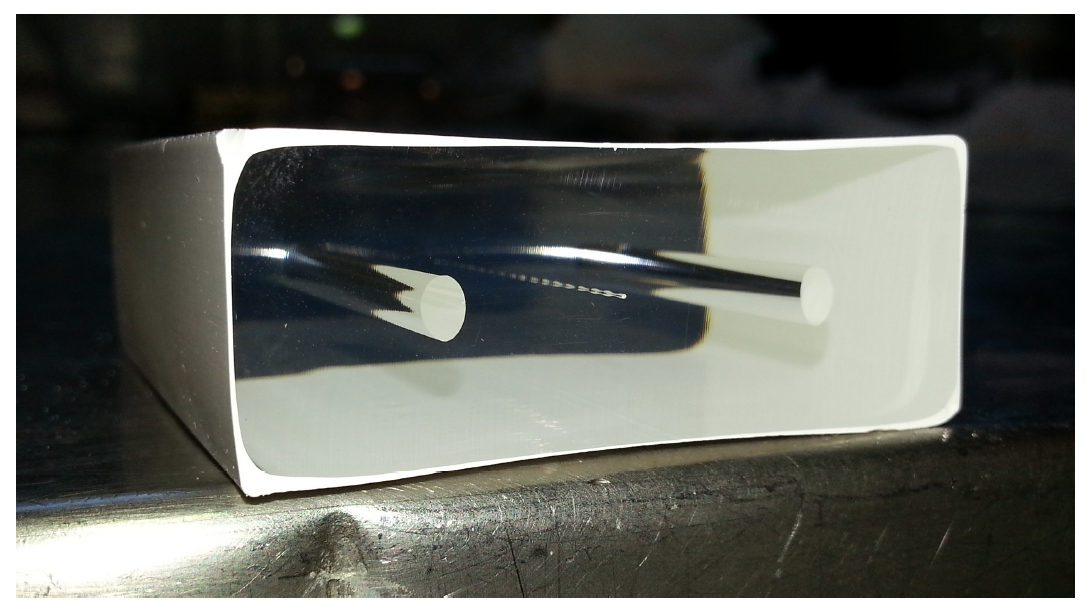}
\caption{[Left]: Detector operating principle, [Right]: Photograph of raw extrusion.} 
\label{fig:prince}
\end{figure}

Plastic scintillator detectors have been used in high-energy physics experiments for decades, and with the development of extruded plastic scintillator at Fermilab~\cite{MINOS:1998kez,PlaDalmau:2003if,PlaDalmau:2000bf,Michael:2008bc,MINERvA:2013zvz,Ambrosi:2011zz,Ambrosino:2014kra,Anastasio:2013sva}, their use has expanded considerably. 
A recent example of an extruded scintillator detector is the one that has been developed for the Mu2e experiment at Fermilab~\cite{Donghia:2018duf}. This experiment requires approximately 1200 m$^2$ of a very efficient detector for cosmic-ray muons. We believe that this concept can be effectively extended to the radiation detection applications described in this paper.
The active element is the ``Di-counter unit'' (Figure~\ref{fig:dicounter} [Left]) which consists of two extruded polystyrene-based scintillator strips, each with two holes for wavelength-shifting (WLS) fibers. The strips have a co-extruded titanium dioxide cladding which acts as a reflector. The mechanical specifications for the Di-counter are given in Figure~\ref{fig:dicounter} [Right].
A basic detector panel would consist of 3 ``Di-counters.''  The length of the Di-counter will depend on the application.
When a charged particle, such as a muon, passes through the scintillator, it loses energy that is then converted into blue light by the scintillator.  Similarly, if an X-ray interacts in the plastic, the scattered Compton electron produces light in the scintillator.  Some of this blue light is absorbed by the WLS fiber and is shifted into the green, where roughly 5\% (in each direction) of the green-shifted light is piped along the fiber to a photo-detector~\cite{allan1973,Ghatak_Thyagarajan_1998,site:Kuraray}. By using the WLS fibers to guide the light to the photo-detectors, it is possible to make large detectors with good light collection efficiency even when the particle hits far from the photo-detector.~\footnote{In the Mu2e experiment, the longest strips are 6.9~m.} Figure~\ref{fig:prince} [Left] gives a schematic of this operating principle while Figure~\ref{fig:prince} [Right] shows a photograph of a sample of the raw extrusion.

\begin{figure}[b]
\centering
\includegraphics[width=0.75\columnwidth]{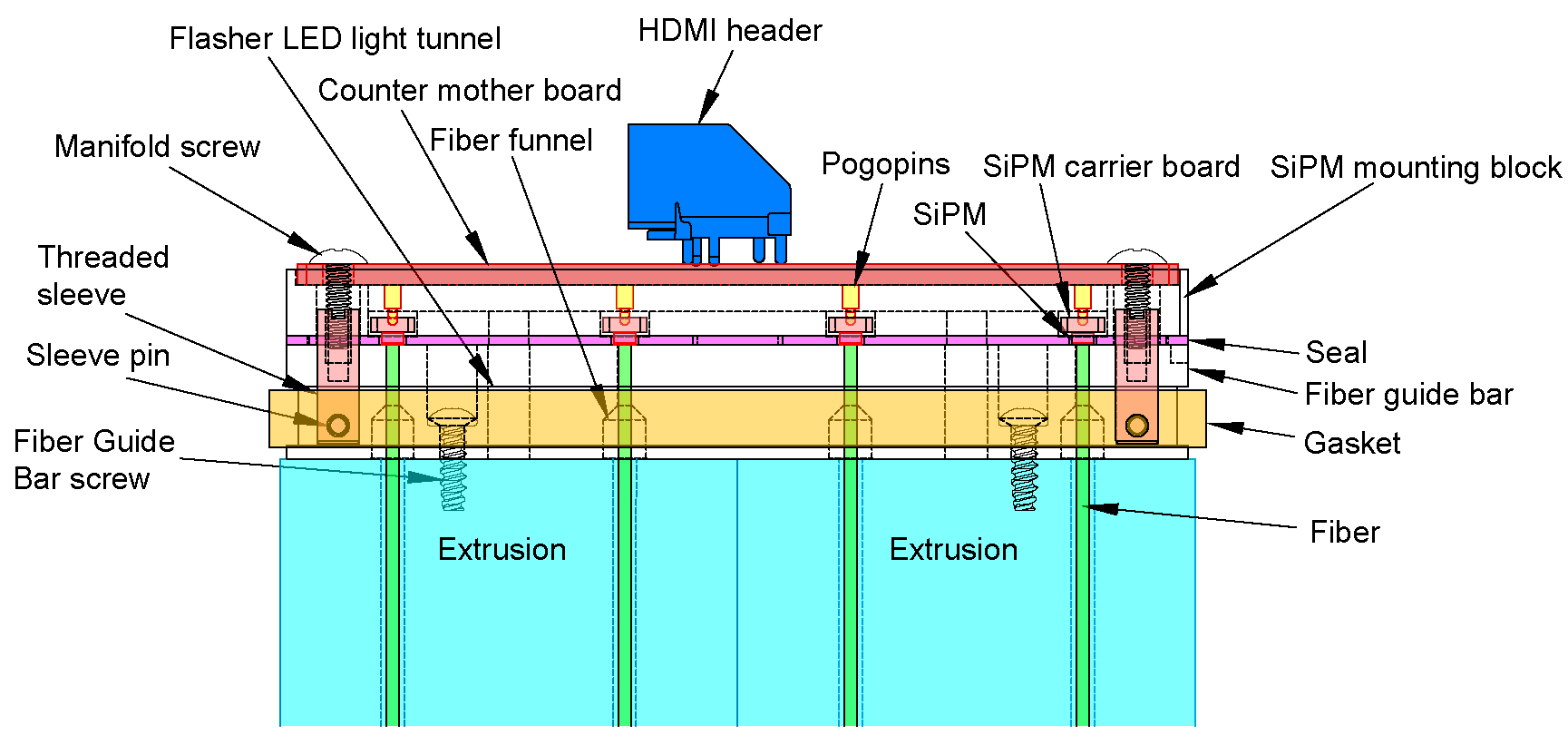}
\caption{Schematic of the Di-counter assembly.} 
\label{fig:Extru}
\end{figure}

The extrusions used for this work were fabricated in the FNAL-NICADD Extrusion Line Facility at Fermilab following the methodologies we have developed over the past 20 years.  Figure~\ref{fig:Extru} shows a schematic of the full Di-counter assembly. 
%
The tests described in this paper used a Di-counter that was assembled with 1.4 mm diameter WLS fiber.  In addition, for this sample, glue (EJ500)~\cite{BC600} was used to fill the channel in the extrusion that holds the WLS fiber.  This improves the optical coupling between the bulk scintillation in the extrusion and the WLS fiber.  
The photo-detectors we use are silicon photomultipliers (SiPMs)~\cite{Beischer:2011zz}.

\begin{figure}[h]
\centering 
\includegraphics[width=0.60\columnwidth]{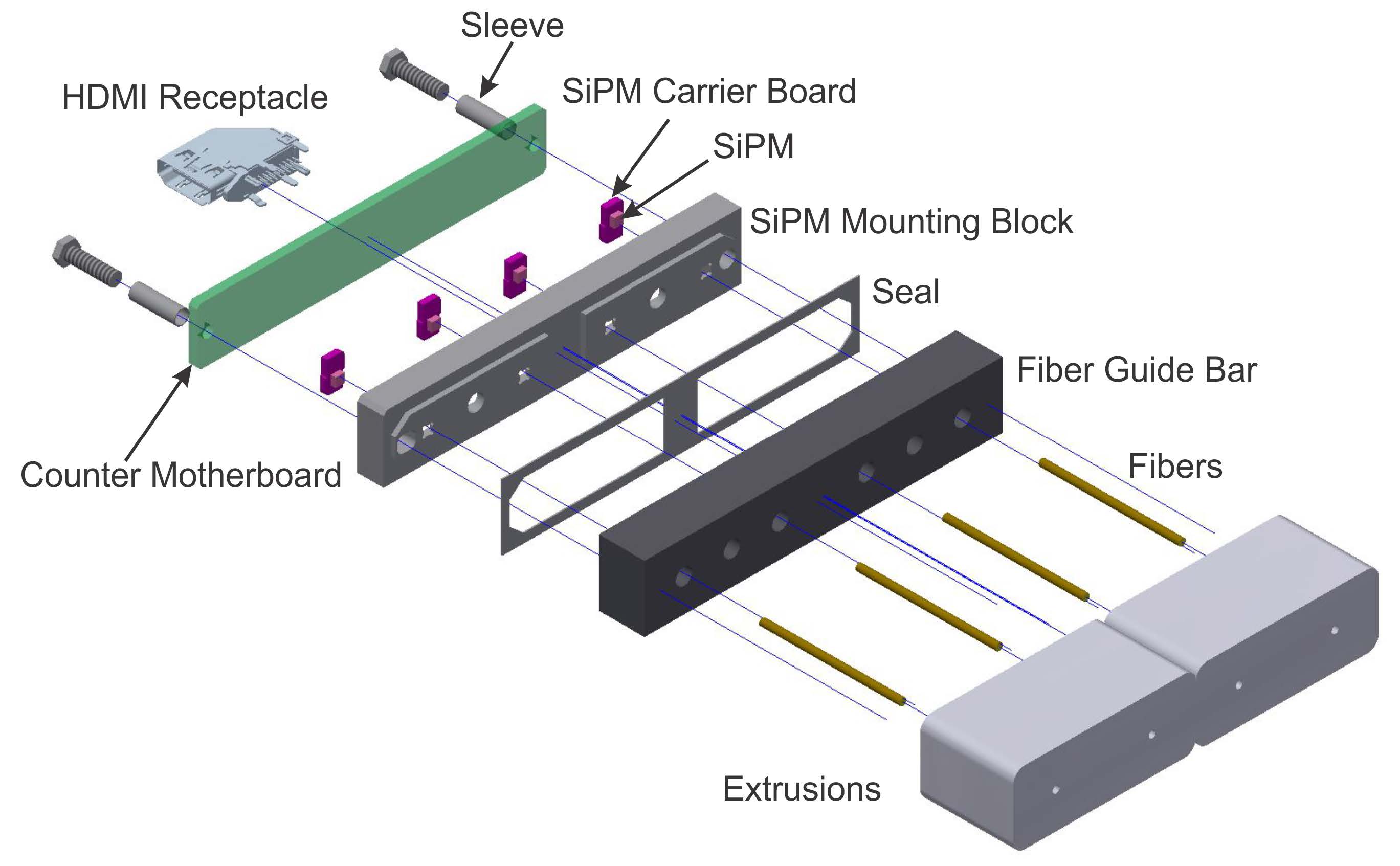} 
\caption{Exploded view of Di-counter readout module.} 
\label{fig:explod}
\end{figure}

A novel mechanical system was designed to align the fibers with the SiPMs, which are carried on a consumer-grade PC board. 
The photo-detector system is modular in nature.  An exploded view of the components is given in Figure~\ref{fig:explod}.  There are 4 SiPMs in each photo-detector module with a module being mounted on each end of the Di-counter so that both ends of the WLS fiber are read out.
The SiPM module carries 4 SiPMs that are 2 $\times$ 2 mm$^2$.

\begin{figure}[b]
\centering
\includegraphics[width=0.60\columnwidth]{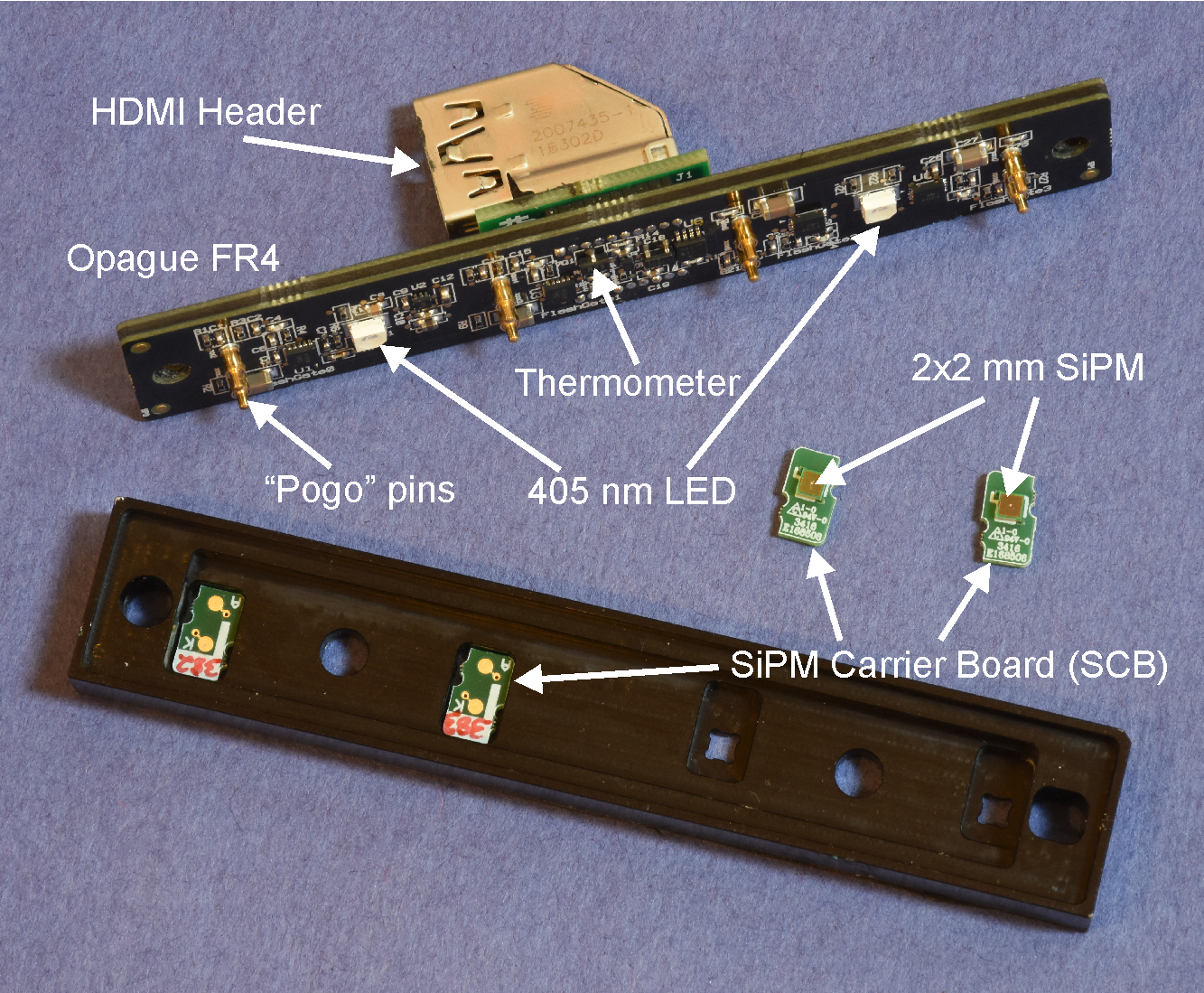}
\caption{Photograph of the SiPM module showing the square SiPMs mounted on their carrier PC board, the readout PC board with HDMI connector, electrical connectors to SiPMs (Pogo pins) and electronics.}
\label{fig:SiPM_mod}
\end{figure}

Figure~\ref{fig:SiPM_mod} is a photograph of the module components.  The SiPM modules can easily be removed from the scintillator extrusions so that they can be characterized separately.  This allows the user to accurately set thresholds to remove the SiPM noise.
A readout system that uses commercial off-the-shelf parts was used in the tests reported in this paper~\cite{Artikov:2017lsc}.

We see two immediate applications for this technology.  The first application is for
food safety monitors.  In areas where radioactive contamination of food products (primarily seafood) is a problem, we shall demonstrate that this technology can provide a high-sensitivity system that can flag unsafe seafood.  It will provide an ``in situ'' background-subtracted counting environment that can quickly (within 10 sec.) flag unsafe food.  Backgrounds from cosmic-ray interactions are easily rejected due to their very large charge deposition and subsequent large electronic signal.  A new approach using triangular extrusions~\cite{Bross:2022ffy} adds the capability to reject cosmic-ray muons using event topology, further improving cosmic-ray rejection.
The second application is for a
radiation detection portal or urban-area radiation monitors. These require a high sensitivity to a number of radionuclides such as, $^{109}$Cd, $^{57}$Co, and $^{137}$Cs.  Neutron sensitivity can be provided with a specialized extrusion structure (see Section~\ref{sec:Neutron}).
%
%
\section{Test results}
\label{sec:PR}
Section~\ref{sec:Cal137} describes the calibration of the detectors,
in Section~\ref{sec:Source} the detector's performance for the detection of a radioactive 
source is reported, and Section~\ref{sec:Food} presents data on the system's capabilities for flagging food contamination.
\subsection{Calibration with a $^{{\bf 137}}$Cs  source}
\label{sec:Cal137}
\begin{figure}[h!]
\centering 
\includegraphics[width=0.99\columnwidth]{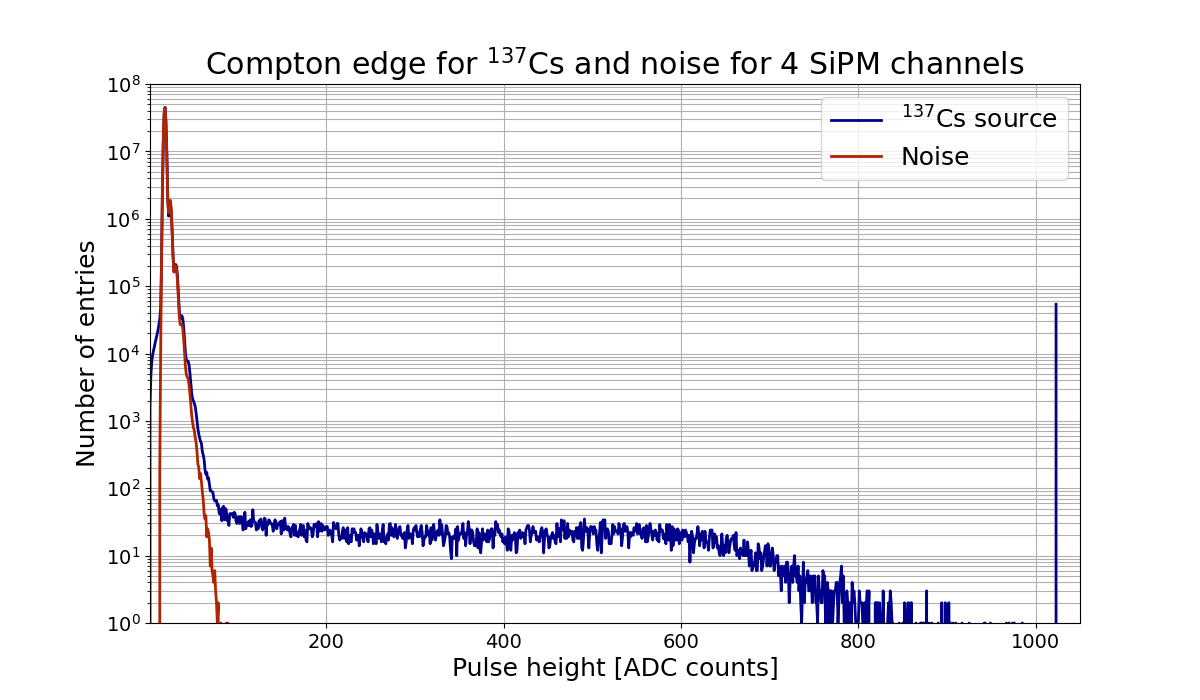}
\caption{Noise histogram for sum of 4 SiPMs on one Di-counter and $^{137}$Cs signal (source distance = 0) from the sum of the 4 SiPMs on one Di-counter.} 
\label{fig:noise}
\end{figure}

Because the detector is modular (photo-detector module separate from scintillator detector), determining a threshold cut in ADC counts to remove the SiPM noise is very straightforward.  The SiPM modules are removed from the scintillator and placed in a light-tight enclosure, and data are then taken.  An integration time of 10 s is used to produce histograms of the SiPM noise distribution.  Figure~\ref{fig:noise} shows one such histogram (Noise) for one counter in the Di-counter (sum of four SiPM signals - 2 WLS fibers with double-ended fiber readout).  In order to effectively eliminate a large contribution from the SiPM noise, a cut in ADC counts is chosen to produce a summed SiPM noise rate of $\simeq$ 2 Hz.

The energy calibration is determined from the position of a $^{137}$Cs Compton edge in ADC counts.  A 2.22$\times10^4$ Bq $^{137}$Cs source is placed in direct contact at the midpoint of one of the counters in the Di-counter.  An integration over 10 s is performed and the sum of the output from the 4 SiPMs is measured.  The system gain is set so that the Compton edge is at $\simeq$ 500 ADC counts.  A typical distribution is also shown in Figure~\ref{fig:noise} ($^{137}$Cs source).
The Compton edge gives the equivalent energy deposition per ADC count since we know that the energy of the Compton edge can be determined from:
\begin{equation}
    E_{\textrm {Edge}} = E \left( 1-\frac{1}{1+\frac{2E}{m_ec^2}} \right),
\end{equation}
where $E$ is the $^{137}$Cs photo-peak energy (662 keV).  The Compton edge is thus at $\simeq$ 480 keV.  

From Figure~\ref{fig:noise}, we see that the Compton edge is at $\simeq$ bin 515.  The pedestal is at bin 15, so the energy per bin is  $\simeq$ 1 keV.   Note: The overflow ADC count (1024) shown in Figure~\ref{fig:noise} captures cosmic-ray muon events.

%
\subsection{Data with the $^{{\bf 137}}$Cs source}
\label{sec:Source}
\begin{figure}[h]
\centering 
\includegraphics[width=0.99\columnwidth]{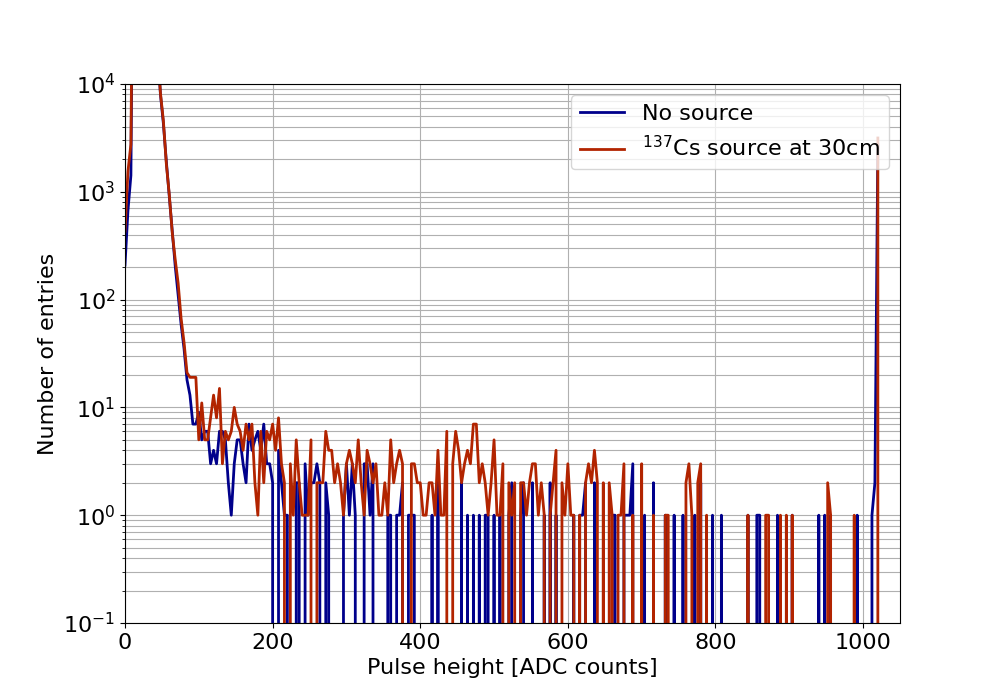}
\caption{A 10 s integration of the signal from one counter with no source present (blue) overlayed with a 10 s integration of the signal from one counter with the source 30 cm from the counter (red).} 
\label{fig:bkg}
\end{figure}

The data described in this section used the same 2.22$\times10^4$ Bq $^{137}$Cs source that was used for the calibration described in Section~\ref{sec:Cal137} with readout of one of the counters in the Di-counter (all 4 fiber ends).  First data are taken without the source in place and then data are taken with the source positioned 30 cm above one of the counters in the Di-counter.  Data taken without the source give us the terrestrial background rate in our laboratory.
%
%
To measure the flux from the $2.22\times10^4$ Bq $^{137}$Cs source, an ADC cut at 77 counts yielded the best results.  With this ADC cut, the count was 391 for a 10 s ($\simeq$ 39 Hz) integration with no source in place; the data are shown in blue in  Figure~\ref{fig:bkg}.  The data shown in red in Figure~\ref{fig:bkg} are with the $^{137}$Cs source positioned 30 cm above the counter.  In this case, the count is 696 for the 10 s integration or $\simeq$ 70 Hz.\footnote{Note: In the hit sum for both background and $^{137}$Cs, the content of the ADC overflow counts (1024) is not included.}
The differential count rate (with source-without source) was $\simeq$~31 Hz.  Using the source activity ($2.22\times10^4$ Bq), the source distance of 30 cm, the counter area and thickness (30 cm $\times$ 5 cm $\times$ 2 cm thick) and the stopping power of polystyrene at 662 keV, for full efficiency we would expect a count rate of $\simeq$ 53 Hz. Therefore, this counter is $\simeq$ 60\% efficient for the detection of $^{137}$Cs photons that interact in the scintillator.  As a cross check, for the data with no source, we integrated the flux above ADC channel 77, weighted by 1 keV per ADC count to obtain a total integrated exposure for the 10 s.  Assuming a 60\% efficiency for the background photons also and extrapolating to a 1 year exposure, we obtain $\simeq$ 30 mrem, which is consistent with the terrestrial background.  Cosmic-ray counts (overflow count) are not included and account for $\simeq$ half the total yearly exposure at sea level.
We have tested this counter's performance against a commercial standard, a Bicron Analyst~\cite{site:Bicron}.  Table~\ref{tab:Comp} reports on our results.  In each case, the system's sensitivity to our $^{137}$Cs source was measured. The integration time (10 s) and the distance (30 cm) were fixed.  In the table, S+Bkg is the integrated number of counts with the source in place, Bkg the integrated number of counts with no source, $\sigma$ represents the $\sqrt{\mathrm{Bkg}}$, and S$_B$ is the number of $\sigma$ the signal is above background.  In all cases, the sum starts at ADC channel 77 and excludes channel 1024.
As can be seen in Table~\ref{tab:Comp}, our panel outperforms the commercial standard which uses a NaI crystal for its detector.
\begin{table}[h]
    \centering
    \caption{Comparison between Di-counter and Bicron Analyst for $^{137}$Cs source excitation.}
\begin{tabular}{|c|c|c|c|c|c|c|c|}
\hline
Device & T(s) & Dist.(cm) & S+Bkg & Bkg & S-Bkg(Hz) & $\sigma$ & S$_B$\\
\hline
$\frac{1}{2}$ Di-counter & 10 & 30 & 696 & 391 & 30.5 & 19.8 & 15\\
\hline
Bicron Analyst & 10 & 30 & 327 & 217 & 11 & 15 & 7\\
\hline
\end{tabular}
\label{tab:Comp}
\end{table}
%
%
\subsection{Food safety monitoring test}
\label{sec:Food}
In order to study food safety applications for our detector, we have used our prototype panel to measure the count rate for 100 grams of Brazil nuts,
which have a nominal activity~\cite{site:ORAU} of 10 Bq.~\footnote{Food safety standards in the US use a limit of 1000 Bq/kg, but in Japan~\cite{FS_japan}, in part in reaction to illegal fishing after the Fukushima accident, the standards are stricter, and food is deemed safe if the activity is less than 100 Bq/kg.}  In our tests, we evenly spread the 100 grams of Brazil nuts directly on top of one counter of a Di-counter.  The measured signal above background due to the activity of the nuts was $\simeq$ 3 Hz. In this study, we found that the optimal S:N ratio was achieved with an ADC cut of 190, which is significantly higher than that needed to reduce the SiPM noise to 2 Hz.  The Di-Counter data in Table~\ref{tab:nuts} were obtained using this ADC cut and rejection of the overflow bin. We also used the Bicron Analyst to detect the radiation from the nuts by surrounding its detector with the nuts.  In the table, $\sigma$ is the square root of the ``Food off" value.
\begin{table}[h!]
    \centering
    \caption{Sensitivity to the radioactivity from Brazil nuts for a 10 s exposure.  ``Food on" is the count with the Brazil nuts on the scintillator, ``Food off" is the background count and $\sigma$ is the square root of the ``Food off" value.}
\begin{tabular}{|c|c|c|c|c|c|}
\hline
Device & T(s) & ``Food on" & ``Food off" & $\sigma$ & \verb|#| $\sigma$ above ``Food off"\\
\hline
$\frac{1}{2}$ Di-Counter & 10 & 102 & 70 & 8.4 & 3.8\\
\hline
Bicron Analyst & 10 & 137 & 131 & 11.4 & 0.5\\
\hline
\end{tabular}
\label{tab:nuts}
\end{table}

In this example, our panel performs much better than the NaI-based detector.  This is due to the large surface area (relative to the Bicron Analyst's NaI crystal) of our counter, which results in much better effective stopping efficiency for the radiation.  In a practical installation at point-of-purchase, the sensitive area of the radiation panel would be roughly 30 cm $\times$ 30 cm or roughly 6 times the area of the single counter of the Di-counter used in the test described above.  The background rate would, therefore, increase by 6, reaching $\simeq$ 420 counts in our 10 s integration window.  One standard deviation above background would be $\simeq$ 20 counts.  For a 1 kg sample with 100 Bq of radioactive contamination placed on this 30 cm $\times$ 30 cm counter, and based on the results given above, we would have a count above background of 300 counts (30 Hz $\times$ 10 sec.) or $\simeq$ 15$\sigma$ above background.  This case clearly demonstrates the efficacy of our technology for flagging, with very high efficiency and very low false-positive rate, food borne radioactive contamination at the 100 Bq/kg level.
\section{National security application}
\label{sec:ManH}
In the course of the study reported in Section~\ref{sec:PR}, we have also evaluated how well our technology will work for detecting illicit nuclear material.  
The US Department of Homeland Security has proposed a radiation portal specification~\cite{PNNL-TM,10108546} for $^{137}$Cs. The specification requires a net (above background) count rate of 110 Hz for $1 \mu$Ci ($3.7\times10^4$ Bq) for a detector that is 2 m from the source.  Additional specifications ask for a minimum of 6100 cm$^2$ of detector area whose aspect ratio (length/width) is a minimum of 4.  For this analysis we use two stations (located $\pm$ 2 m from the source) each 2 m tall $\times$ 50 cm wide with two layers deep of Di-counter of the geometry shown in Figure~\ref{fig:dicounter}.  The total depth is therefore 4 cm.   In this analysis we have followed the Radiation Portal Monitors (RPMs) concept shown in Figure 1-1 of~\cite{PNNL-TM} and reproduced in Figure~\ref{fig:portal} below.
\clearpage

We now extrapolate the panel performance described in Section~\ref{sec:Source} to the parameters given above by scaling by surface area and by stopping power (thickness) and using a $ 1/r^2$ dependence on the source to detector distance.  The scaling is shown in Equation~\ref{eq:Rate} below.  The first term corresponds to the source strength used in Section~\ref{sec:Source} relative to the 1 $\mu$Ci specification and the second term corresponds to the signal rate in Hz from Section~\ref{sec:Source}.  We then multiply by the ratio of the detector areas, the number of detectors (2), the relative detector thickness (2) and finally the 1/$r^2$ dependence on the source distance.  The final signal over background rate, R, is:
\begin{equation}
    \mathrm{R}=\left(\frac{1}{0.6}\right)\times30\times\left(\frac{200\times50}{150}\right)\times2\times2\times\left(\frac{30}{200}\right)^2 \simeq 300~ \text{Hz.}
    \label{eq:Rate}
\end{equation}
Our proposed portal configuration greatly exceeds the specification.  The signal to noise will depend on the location of the portal, but from the results is Section~\ref{sec:Source} we expect the signal to be $\simeq 3-4\sigma$ over background.
\begin{figure}[t!]
\centering 
\includegraphics[width=0.7\columnwidth]{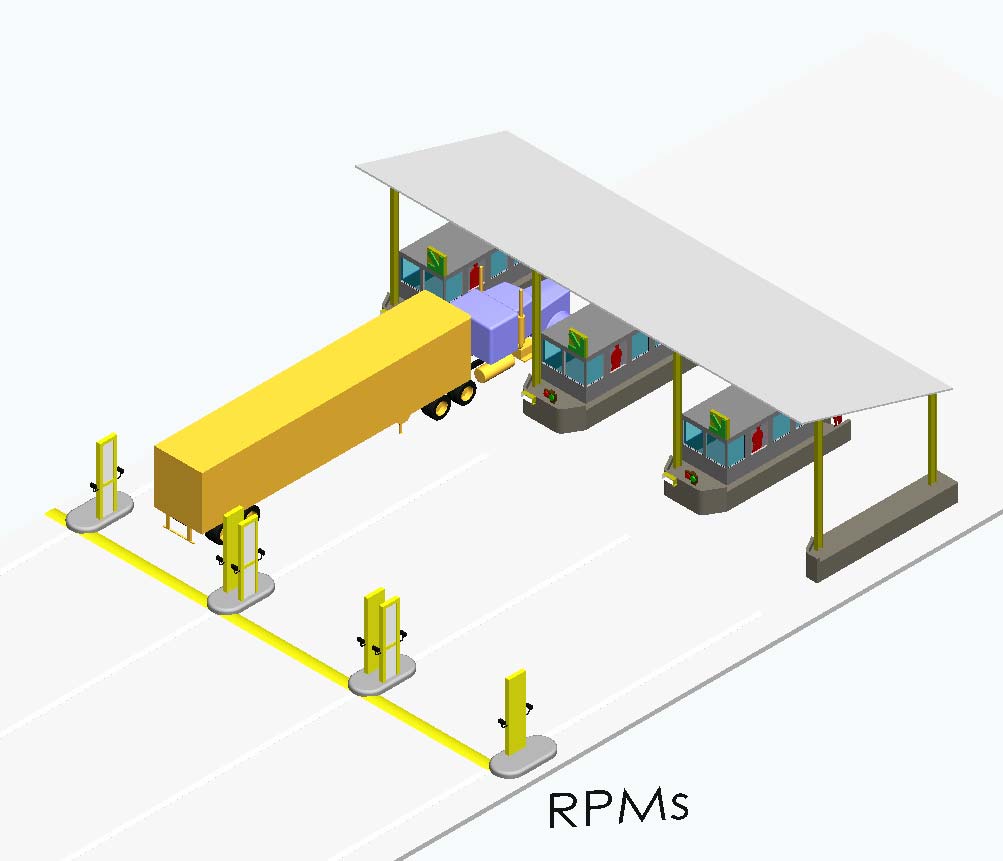}
\caption{General Portal Monitor System operational concept.} 
\label{fig:portal}
\end{figure}

We next consider the effectiveness of this technology to protect large areas from illicit radiation sources, such as might be used in a "dirty bomb"~\cite{DirtyB} attack by terrorists. Here we give an idea of how an array of counters {\it (Radiation Cell Towers)} with the type of performance we demonstrated above for $^{137}$Cs could be used to cover a city.  As an example, we use the NY borough of Manhattan.  Manhattan has a total area of $\simeq$ 60 km$^2$. We assume the performance level indicated above and extend the area of each panel to 60 cm $\times$ 30 cm (still small and easily deployable on light poles, for example).  The metric we use here is the detection of the equivalent of an unshielded $3.7\times10^8$ Bq $^{137}$Cs source.  This would be equivalent to approximately 6 cm of lead surrounding a $3.7\times10^{14}$ Bq source.  Table~\ref{tab:Manh} summarizes the performance of this 60 cm $\times$ 30 cm detector panel extrapolated from the measured detector performance described in
Section~\ref{sec:Source}.
%
%
A single panel can detect the $3.7\times10^8$ Bq from an un-shielded $^{137}$Cs source at a distance of 245~m, assuming a ten second integration window and a $3\sigma$ above background threshold. A single panel can cover an area of approximately 0.084 km$^2$ based on this number.  
Therefore, $\simeq$ 320 such panels could monitor the entire borough of Manhattan.  An estimated system cost is given in the Appendix.  In this example, we used the background rate observed in our laboratory.  In the field mounted on light poles, the background could be less due to the detectors' being located away from concrete, asphalt and dirt.
\begin{table}[h]
    \centering
    \caption{Performance of Radiation Cell Tower counter.}
\begin{tabular}{|c|c|c|c|c|}
\hline
Area & T(s) &  Bkg & $\sigma$ = $\sqrt{\textrm {Bkg}}$ & Dist.(m) when signal is 3$\sigma$ above Bkg.\\
\hline
60x30 cm$^2$ & 10 & 4700 & 69 & 245\\
\hline
\end{tabular}
\label{tab:Manh}
\end{table}
\section{Neutron sensitivity enhancement}
\label{sec:Neutron}
Although scintillator detectors have sensitivity to fast neutrons, it is difficult to distinguish a neutron event from ordinary ionizing radiation.  In addition, conventional plastic scintillator has limited capability to distinguish a neutron interaction from ionizing radiation. The extruded plastic scintillator technology that has been developed at Fermilab can be extended to produce a neutron sensitive plastic and, if there is an argument for including neutron detection capability such as to aid detection of weapons grade plutonium, a second detector could be added to our system to include this function~\cite{Neutron_TN,bross2005method,bross2006method}.  

An extruded scintillator that is only sensitive to thermal neutrons is based on the concept shown in Figure~\ref{fig:NSS} and uses the following reaction:
\begin{figure}[b]
\centering 
\includegraphics[width=0.8\textwidth]{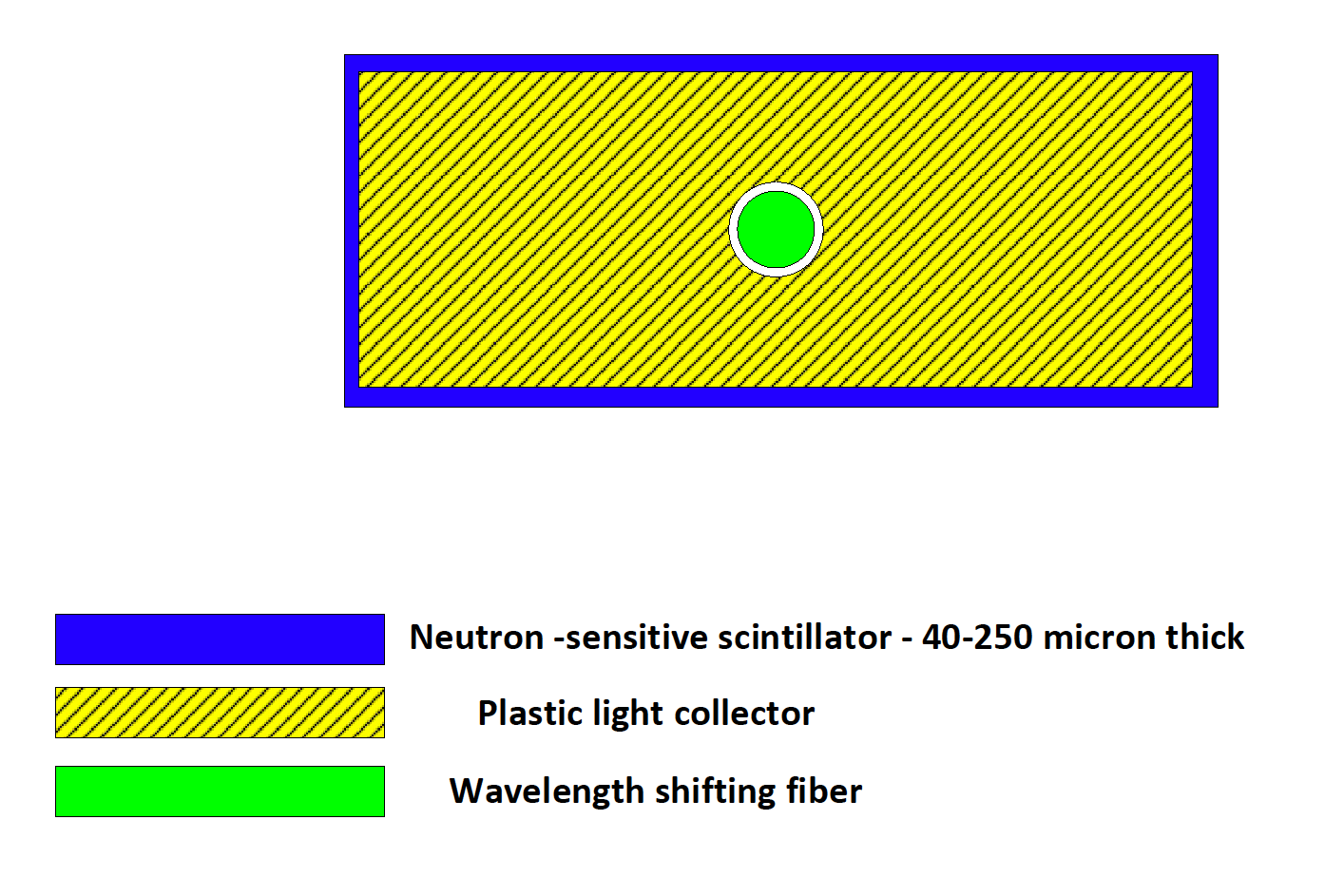} 
\caption{Concept for an extruded scintillator that would be preferentially sensitive to slow neutrons.} 
\label{fig:NSS}
\end{figure}
\begin{center}
n + $^6$Li $\rightarrow$ $^4$He + $^3$H + 4.79 MeV.
\end{center}
The neutron-sensitive scintillator layer shown in Figure~\ref{fig:NSS} is made by introducing $^6$LiF nano-particles into a conventional polystyrene-based scintillator during the extrusion process.  The daughter particles in the reaction (alpha + triton) deposit all their energy in a very thin layer ($\simeq 50~\mu$m).  In this way, a thin active layer can efficiently detect neutrons, where gammas and minimum ionizing particles would deposit very little energy and their signal would fall below threshold.  The thin layer would be co-extruded onto undoped polystyrene which acts as a light collector bringing the light to a WSF for readout as described above.  In addition, an outer reflector cladding could be added to improve the light yield.  Our extrusion process lends itself in a very natural way to producing this detector.  

\begin{figure}[h!]
\centering 
\includegraphics[width=0.99\textwidth]{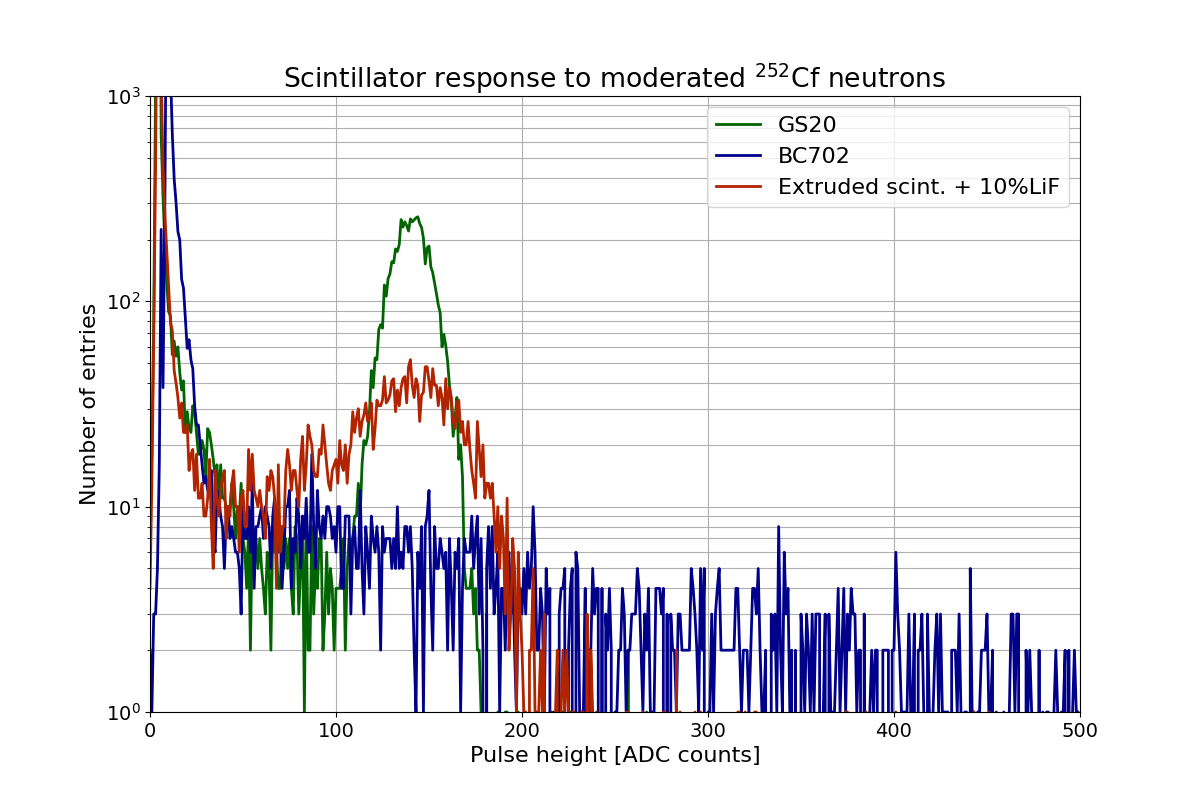} 
\caption{Pulse height spectra for neutron scintillators.} 
\label{fig:NData}
\end{figure}

To date we have not had the equipment to produce the extrusion shown in Figure~\ref{fig:NSS}, but we have produced $^6$LiF loaded polystyrene scintillator and compared its light yield for thermalized neutrons from a $^{252}$Ca source to two common solid neutron scintillator standards: GS20~\cite{GS20}  (neutron sensitive glass scintillator) and BC702~\cite{BC720}  (neutron sensitive ZnS(Ag)-$^6$LiF scintillator).  A 4” lead shield was used to stop gammas for these measurements.  Disks of each scintillator ($^6$Li loaded polystyrene 30 mm diameter x 1mm thick - 2.5 mg $^6$Li/cm$^2$, GS20: 25 mm diameter x 2 mm thick – 17.2 mg $^6$Li/cm$^2$, BC702: 50 mm diameter – 11 mg $^6$Li/cm$^2$) were placed directly on a Hamamatsu R669 PMT~\cite{R669} and the PMT was connected to a LeCroy qVt~\cite{qVt} pulse height analyzer.  The pulse height distributions for these three scintillators are given in Figure~\ref{fig:NData}.  The light yield of the $^6$LiF loaded plastic compares quite favorably.
%
%
\section{Conclusions}
We have shown in this paper that modern extruded plastic-scintillator technology, using solid-state light detectors for detection of the scintillation light, can find far-reaching use in food safety and national security applications.  For food safety, we have shown that our technology can detect 100 Bq/kg.  For radiation detection in security applications, our system has a sensitivity of $\simeq$ 3 times the US Department of Homeland Security's minimum specification.  There is still room for increased performance levels through directed detector R\&D and optimization, but most of the underlying technology base is firmly in place.  The cost--performance envelope for these types of systems is very attractive.
\appendix
\section{Cost model for Radiation Cell Tower system}
\label{appen}
A rough cost estimate for the city monitoring system described in Section~\ref{sec:ManH} is given here.
To fully engineer an environmentally robust detector would require $\simeq$ \$10M in non-recurring engineering (NRE) costs, which dominates the cost for the first system of this type.  Once in production, we estimate a unit cost of $\simeq$ \$2,000 with installation and infrastructure costs of \$3,000 per unit.  Total investment is then on the order of \$12M for the first system including the NRE.  A detailed cost model is given in Table~\ref{tab:Manh_c}.
%
%
%
\begin{table}[h!]
    \centering
    \caption{Breakdown of estimated costs for the city monitoring system described in Section~\ref{sec:ManH} including non-recurring engineering (NRE), component costs, assembly, test, and installation.}
\begin{tabular}{|c|c|c|c|}
\hline
{\bf Component} &  {\bf Number} &  {\bf Cost (\$)} & {\bf Total (k\$)}\\
\hline
Mechanical engineering (NRE) & Lot & - & 3000\\
\hline
Electrical engineering (NRE) & Lot & - & 5000\\
\hline
Software engineering (NRE) & Lot & - & 2000\\
\hline
Scintillator & 320 & 100 & 32\\
\hline
Fiber & 7680 & 3 &  23\\
\hline
Photodetector module & 7680 & 5 & 38\\
\hline
Electronics & 7680 & 20 & 154\\
\hline
Enclosure & 320 & 500 & 160\\
\hline
Assembly and test & 320 & 1000 & 320\\
\hline
Installation & 320 & 3000 & 960\\
\hline
\hline
{\bf TOTAL} & & & {\bf 11687}\\
\hline
\hline
\end{tabular}
\label{tab:Manh_c}
\end{table}
%
\bibliographystyle{JHEP}
\bibliography{FishD}

\end{document}